\documentstyle[12pt,aaspp4]{article}
\begin{document}

\title{The Energy Dissipation Rate of Supersonic, Magnetohydrodynamic
Turbulence in Molecular Clouds}

\author{Mordecai-Mark Mac Low}
\affil{Max-Planck-Institut f\"ur Astronomie, K\"onigstuhl 17, D-69117
Heidelberg, Germany \\E-mail: mordecai@mpia-hd.mpg.de}


\begin{abstract}
Molecular clouds have broad linewidths suggesting
turbulent supersonic motions in the clouds.  These motions are usually
invoked to explain why molecular clouds take much longer than a
free-fall time to form stars.  It has classically been thought that
supersonic hydrodynamical turbulence would dissipate its energy
quickly, but that the introduction of strong magnetic fields could
maintain these motions.  In a previous paper it has been shown,
however, that isothermal, compressible, MHD and hydrodynamical
turbulence decay at virtually the same rate, requiring that constant
driving occur to maintain the observed turbulence.  In this paper
direct numerical computations of uniformly driven turbulence with the
ZEUS astrophysical MHD code are used to derive the absolute value of
energy dissipation, which is found to be
\begin{displaymath}
\dot{E}_{\rm kin} \simeq - \eta_v m \tilde{k} v_{\rm rms}^3,
\end{displaymath}
with $\eta_v = 0.21/\pi$, where $v_{\rm rms}$ is the root-mean-square
velocity in the region, $E_{\rm kin}$ is the total kinetic energy in
the region, $m$ is the mass of the region, and $\tilde{k}$ is the
driving wavenumber.  The ratio of the formal decay time $E_{\rm
kin}/\dot{E}_{\rm kin}$ of turbulence to the free-fall time of the gas
can then be shown to be
\begin{displaymath}
\tau(\kappa)  =  \frac{\kappa}{M_{\rm rms}} \frac{1}{4 \pi \eta_v},
\end{displaymath}
where $M_{\rm rms}$ is the rms Mach number, and $\kappa$ is the ratio
of the driving wavelength to the Jeans wavelength.  It is likely that
$\kappa < 1$ is required for turbulence to support gas against
gravitational collapse, so the decay time will probably always be 
far less than the free-fall time in molecular clouds, again
showing that turbulence there must be constantly and strongly driven.
Finally, the typical decay time constant of the turbulence can be
shown to be
\begin{displaymath}
t_0 \simeq 1.0 \,\,{\cal L} / v_{\rm rms},
\end{displaymath}
where ${\cal L}$ is the driving wavelength.

\end{abstract}


\clearpage
\section{Introduction}

Star-forming molecular clouds appear to have lifetimes more than an
order of magnitude longer than it would take them to gravitationally
collapse in the absence of any support (Blitz \& Shu 1980).  Typical
lifetimes are of order 30 Myr, while the free-fall time 
\begin{equation} \label{tff}
t_{\rm ff} = (3\pi / 32 G \rho)^{1/2} = (1.2 \times 10^6 \mbox{ yr}) (n/10^3
\mbox{ cm}^{-3})^{-1/2},
\end{equation}
where n is the number density of the cloud and I assume the mean
molecular mass $\mu = 3.32 \times 10^{-24}$~g.  The gas in molecular
clouds also appears to be moving in random directions at supersonic
velocities, in a fashion usually described as turbulent.  Evidence for
this includes molecular emission lines an order of magnitude broader
than the thermal linewidth, and the transient, clumpy nature of
the clouds (Blitz 1993).  

These clumps have been studied in some detail by Stutzki \& G\"usten
(1990), and by Williams, De Geus \& Blitz (1994), who showed that
regions of enhanced density can be separated out based on their
coherent velocity structure.  Studies of the formation of
photodissociation regions by penetration of UV radiation into the
clouds independently lead to the conclusion that the gas is extremely
clumpy (Stutzki et al.\ 1988).  Another independent piece of evidence
for extreme clumpiness is the studies of dust extinction of background
stars through molecular clouds by Lada et al.\ (1994) and Alves et
al.\ (1998) that show greater variance in extinction in regions with
greater average extinction.  Models of the chemical abundances in the
dense clumps show that they have lifetimes of only a million years or
less, much shorter than the lifetimes of the clouds as a whole, but
comparable to dynamical and collapse times (Prasad et al. 1991, Bergin
et al. 1997).

It has been clear since the discovery of these supersonic motions that
supersonic turbulence would decay quickly (e.g.\ Field 1978), although
the common argument that it would decay more quickly than subsonic
turbulence due to the extremely dissipative nature of shocks has
turned out not to be correct (Mac Low et al.\ 1998a; hereafter Paper
I): rather, supersonic turbulence decays somewhat more slowly than
subsonic, incompressible turbulence, though both decay quickly.
Magnetic fields have classically been invoked to maintain the observed
supersonic motions.  The argument has been that the presence of a
strong field would transform dissipative shocks to non-dissipative
linear magnetohydrodynamic (MHD) waves (Arons \& Max 1975, Mouschovias
1975).  However, numerical models suggest that the interaction between
even mildly non-linear Alfv\'en waves inevitably generates a spectrum
of Alfv\'en waves with power reaching down to the dissipation scale,
however that may be determined (Mac Low et al. 1998b).  As a result,
compressible magnetohydrodynamic (MHD) turbulence decays at close
enough to the same rate as hydrodynamic turbulence as to not be
astrophysically distinguishable (Paper I).

Therefore the observed supersonic motions must be driven on timescales
short compared to a dynamical time.  Luckily, there is no shortage of
potential driving mechanisms.  In fact, the problem is not one of
finding a plausible driving mechanism, but rather one of choosing from
the multiple suspects at hand.  

Differential rotation of the galactic disk (Fleck 1981) is attractive
as it should apply even to clouds without active star formation.
Furthermore, support of clouds against collapse by shear could explain
the observation that smaller dwarf galaxies, with lower shear, have
larger star-formation regions (Hunter 1998).  However, the question
arises whether this large-scale driver can actually couple efficiently
down to molecular cloud scales.  Balbus-Hawley instabilities might
play a role here (Balbus \& Hawley 1998).  

Turbulence driven by gravitational collapse has the attractive feature
of being universal: there is no need for any additional outside energy
source, as the supporting turbulence is driven by the collapse process
itself.  Unfortunately, it has been shown by Klessen, Burkert, \& Bate
(1998) not to work for gas dynamics in a periodic domain. The
turbulence dissipates on the same time scale as collapse occurs,
without markedly impeding the collapse.  The influence of magnetic
fields on this problem remains an open question, although the results
of Paper I suggest that they will not be important.

Ionizing radiation (McKee 1989, Bertoldi \& McKee 1997,
V\'azquez-Semadeni, Passot \& Pouquet 1995), winds, and supernovae
from massive stars provide another potential source of energy to
support molecular clouds.   Here the problem may be that they are too
destructive, tending rather to destroy the molecular cloud they act on
rather than merely stirring it up.  If the clouds can be coupled to a
larger-scale interstellar turbulence driven by massive stars, however,
perhaps this problem can be avoided.

A final suspect for the driving mechanism is jets and outflows from
the more ubiquitous low-mass protostars that should naturally form in
any collapsing molecular cloud (McKee 1989, Franco \& Cox 1983, Norman
\& Silk 1980), allowing the attractive possibility of star-formation
being a self-limiting process.  It has recently become clear that
these jets can reach lengths of several parsecs (Bally, Devine, \& Alten
1996), implying total energies of order the stellar accretion energy,
as suggested by Shu et al.\ (1988) on theoretical grounds.  However,
it remains unclear whether space-filling turbulence can be driven by
sticking needles into the molecular clouds.

In this paper I consider the most general question required to begin
distinguishing among these different models:  what is the energy
dissipation rate of turbulence uniformly driven at some specified
wavelength by an arbitrary forcing field?  I consider supersonic
turbulence in the presence of magnetic fields with strengths ranging
from zero up to somewhat above equipartition with the gas motions.  In
order to apply my results directly to molecular clouds, I adopt an
isothermal equation of state.

\section{Computational Technique}

To compute the energy dissipation from uniformly driven turbulence I
use the astrophysical MHD code ZEUS-3D (Stone \& Norman 1992a,b).
This is a second-order code using van Leer (1977) advection that
evolves magnetic fields using a constrained transport method (Evans \&
Hawley 1988) modified by upwinding along shear Alfv\'en
characteristics (Hawley \& Stone 1995), and that resolves shocks using
a Von Neumann type artificial viscosity.  It contains no other explicit
dissipation or resistivity terms, but structures with size approaching
the grid resolution are subject to the usual numerical dissipation.

Using numerical dissipation and artificial viscosity as substitutes
for a model of physical dissipation can be justified if the details of
the behavior at the dissipation scale can be separated from the
larger-scale dynamics of the flow.  This assumption appears to be
valid in the case of incompressible hydrodynamic turbulence (e.g.\
Lesieur 1997).  In Paper I we studied with some care the question of
whether we could also make this assumption in the case of decaying
supersonic flow with or without the presence of magnetic fields.  We
performed resolution studies with grids ranging from $32^3$ to $256^3$
zones, and found that at resolutions greater than $64^3$ zones the
power-law in time at which the energy decayed became almost
independent of the grid resolution.  Because the grid resolution
directly determines the scale at which both the numerical dissipation
and the artificial viscosity act, this suggests that these scales (or
equivalently in the case of the numerical dissipation, its strength)
are not important so long as they are separated from the dynamical
scale.  To further test this assumption, we also compared our results
in the hydrodynamic case to computations of the same problem using a
smoothed particle hydrodynamics (SPH) code, and found that this very
different numerical method again gave very similar answers.

Our resolution studies did reveal that in models with magnetic fields,
convergence occurred more slowly in models with initial magnetic
energy close to equipartition with the kinetic energy.  However, the
decay rate monotonically increased with resolution in those models.
That is, in models with strong magnetic fields, increased resolution
resulted in increased, not decreased, dissipation.  Dissipation in
these models probably occurs due to the dissipation of short
wavelength MHD waves.  Higher resolution may better resolve the
production of these small wavelength waves by the interaction of
non-linear longer wavelength waves with one another.

I perform my computations on a three-dimensional, uniform, Cartesian
grid with side $L = 2$, extending from -1 to 1 with periodic boundary
conditions in every direction, using an isothermal equation of state,
with sound speed chosen to be $c_s = 0.1$.  The initial density and,
in relevant cases, magnetic field are both initialized uniformly on
the grid, with the initial density $\rho_0 = 1$ and the initial field
parallel to the $z$-axis.

To set up a turbulent flow I introduce velocity perturbations in a
fashion perhaps too much inspired by models of incompressible
turbulence.  In those models, a purely solenoidal flow drawn from a
field of Gaussian fluctuations with a power spectrum following a power
law $P(k) \propto k^{-q}$ is set up as a reasonable approximation to
the distribution of vortices typical of incompressible turbulence,
with the index of the power spectrum typically close to the Kolmogorov
(1941) value $q = 5/3$.  In simulations of supersonic turbulence, a
power law with $q = 2$ has been found (e.g. Porter, Pouquet \&
Woodward 1992, 1994).  However, this power spectrum appears to occur
for the simple reason that the Fourier transform of a step function is
$k^{-2}$, and Fourier transforms are additive, so the power spectrum
of a box full of shocks is also going to be close to $k^{-2}$.
Therefore, setting up a flow drawn from a field of Gaussian
fluctuations with $P(k) \propto k^2$, whether only solenoidal (as is
done by Padoan \& Nordlund 1997), or also including compressible modes,
will not be a particularly good approximation to the shock structure
typical of supersonic turbulence.  Nevertheless, Gaussian fluctuations
drawn from a field with power only in a narrow band of wavenumbers
around some value $k$ do offer a very simple approximation to
driving by mechanisms that act on that scale.  Comparing runs with
different $k$ then can give some information on how, for example,
turbulence driven by large-scale shearing motions might differ from
turbulence driven by low mass protostars.

Therefore, I initialize the turbulent flow, as described in Paper I,
with velocity perturbations drawn from a Gaussian random field
determined by its power distribution in Fourier space, following the
usual procedure: for each three-dimensional wavenumber $\vec{k}$ with
$k-1 \leq |\vec{k}| \leq k$ I randomly select an amplitude from a
Gaussian distribution around unity and a phase between zero and
$2\pi$.  I then transform the field back into real space to get a
velocity component in each zone, and multiply by the amplitude
required to get the desired initial root mean square (rms) velocity.
I repeat this for each velocity component independently to get the
full velocity field.  Thus the dimensionless wavenumber $k =
L/\lambda_d$ counts the number of driving wavelengths $\lambda_d$ in
the box.

To drive the turbulence, I then normalize this fixed pattern to
produce a set of perturbations $\delta\vec{\nu}(x,y,z)$, and at every
time step add a velocity field $\delta\vec{v}(x,y,z) = A
\delta\vec{\nu}$ to the velocity $\vec{v}$, with the amplitude $A$ now
chosen to maintain constant kinetic energy input rate $\dot{E}_{\rm
in} = \Delta E / \Delta t$.  For compressible flow with a
time-dependent density distribution, maintaining a constant energy
input rate requires solving a quadratic equation in the amplitude $A$
at each time step. For a grid with $N$ zones on a side, each of volume
$\Delta V$, the equation for $A$ is
\begin{equation}
\Delta E = \frac12 \Delta V \sum_{i,j,k = 1}^{N} \rho_{ijk} A \delta
\vec{\nu}_{ijk} \cdot (\vec{v}_{ijk} + A \delta\vec{\nu}_{ijk}).
\end{equation}
I take the larger root of this equation to get the value of $A$.

These computations have no intrinsic scale.  To convert to
astrophysical units, one must specify mass, length, and time scales or
quantities such as density from which these can be derived.  One
useful set of scales that can be specified is the size of the region
considered $L'$, the mean density $\rho'$, and the sound speed $c_s'$.
As an example, if we choose $L' = 0.5$~pc, $c_s' = 0.2$~km~s$^{-1}$,
and $\rho'_0 = 10^4 (2 m_H)$~g~cm$^{-3}$, then the computational time unit
$t$ can be converted to seconds as 
\begin{equation} \label{tsc}
t' = (L'/L)(c_s/c_s') t = (4 \times 10^{12} \mbox{ sec})
\left(\frac{L'}{0.5 \mbox{pc}}\right) 
\left(\frac{c_s'}{0.2 \mbox{ km s}^{-1}}\right)^{-1} t
\end{equation}
in our example.
Similarly, velocities are scaled with the sound speed
\begin{equation} \label{vsc}
v_{\rm rms}' = (c_s'/c_s) v_{\rm rms} = (2 \mbox{ km s}^{-1})
\left(\frac{c_s'}{0.2 \mbox{ km s}^{-1}}\right) v_{\rm rms},
\end{equation}
energies scale as
\begin{equation} \label{esc}
E' = \frac{\rho'_0}{\rho_0} \left(\frac{L'}{L}\right)^3
\left(\frac{c'_s}{c_s}\right)^2 E = 
(6 \times 10^{44} \mbox{ erg}) 
\left(\frac{L'}{0.5 \mbox{pc}}\right)^3
\left(\frac{c_s'}{0.2 \mbox{ km s}^{-1}}\right)^2
\left(\frac{n'_0}{10^4 \mbox{ g cm}^{-3}}\right) E,
\end{equation}
energy input or dissipation rates scale as 
\begin{equation} \label{edotsc}
\dot{E}' = \frac{\rho'_0}{\rho_0} \left(\frac{L'}{L}\right)^2
\left(\frac{c'_s}{c_s}\right)^3 \dot{E} = 
(4 \times 10^{-2} L_{\odot}) 
\left(\frac{L'}{0.5 \mbox{pc}}\right)^2
\left(\frac{c_s'}{0.2 \mbox{ km s}^{-1}}\right)^3
\left(\frac{n'_0}{10^4 \mbox{ g cm}^{-3}}\right) \dot{E},
\end{equation}
and so forth.


\section{Energy Dissipation}

From dimensional arguments, one expects turbulent energy dissipation
rates $\dot{E_{\rm kin}} = \eta {\cal V}^3/ {\cal L}$, where $\cal V$
and ${\cal L}$ are respectively the characteristic velocity and length
scale of the turbulent region.  However, there are several possible
length and velocity scales available.  The length scale could, for
example, be the size of the box, $L$, or the typical driving
wavelength $\lambda_d$, while the velocity scale could be the sound
speed $c_s$, the Alfv\'en speed $v_A$, or, as found in one dimension
by Gammie \& Ostriker (1996), the current mean turbulent velocity
$v_{\rm rms}$.  There is also no good theoretical derivation of the
value of the constant of proportionality $\eta$ for strongly
compressible turbulence, with or without magnetic fields.  The
numerical simulations described above are designed to determine $\eta$
and to decide which of the potential values of $\cal V$ and $\cal L$
are correct.

Our resolution studies of models of decaying compressible hydrodynamic
and MHD turbulence in Paper~I showed that, for the hydrodynamic cases,
$128^3$ zones captured the decay rate to within a few percent, and
even for the MHD cases, this resolution was good to better than 10\%.
This resolution is also low enough to allow me to do a reasonably
sized parameter study on the machines available to me, so I choose it
for my standard resolution.  I also perform a few runs at $256^3$ to
check the behavior of my results with increasing resolution, however.
In Table~\ref{runs} I describe the runs at standard resolution
discussed in this paper.  The model names begin with either H for
hydrodynamic or M for MHD, then have a letter from A to E specifying
the level of energy input $\dot{E}_{\rm in}$, then a number giving the
dimensionless wavenumber $k$ chosen for driving, and then, for the MHD
models, another number indicating the initial field strength specified
by the ratio of the Alfv\'en speed to the sound speed, $v_A/c_s$.

To compute the equilibrium values of kinetic energy $E_{\rm kin}$ and
root-mean-square (rms) velocity $v_{\rm rms}$, I took time samples
every $2.5 \times 10^{-3} t_s$, where the sound-crossing time $t_s =
L/c_s$.  After waiting $0.2 t_s$ for the turbulence to reach what
appeared from the time history to be an approximate steady-state
equilibrium, I took the remaining points and computed their mean and
variance, typically using several hundred samples.  In a few cases,
the runs were shorter due to the expense of computing with high
Alfv\'en speeds, though conversely equilibrium was reached more
quickly, so I started the averaging at an earlier time to ensure
sufficient samples for a meaningful average.  The reported quantities
have variances under 5\% of the mean, except for the kinetic energies
of the two hydrodynamic models driven with wavenumber $k=2$, which had
large variances as noted in the table.  Driving at large wavenumber in
the absence of magnetic fields produces large structures, whose
interactions introduce larger fluctuations than usual around the mean.

I find that the best description of my models comes by taking a length
scale ${\cal L} = \lambda_d$ and a velocity scale ${\cal V} = v_{\rm
rms}$.  Figure~\ref{diss}{\em (a)} shows equilibrium energy
dissipation rates for all the models in Table~\ref{runs}, compared to
the quantity $k v_{\rm rms}^3 \sim v_{\rm rms}^3/\lambda_d$.  A fit to
the hydrodynamic models HA8 through HE8 gives a relation with slope
1.02.  Let us define a dimensionalized wavenumber $\tilde{k} =
(2\pi/L) k = 2\pi/\lambda_d$.  A very good approximation is then the
linear relation
\begin{equation} \label{vdiss}
\dot{E}_{\rm kin} \simeq -\eta_v m \tilde{k} v_{\rm rms}^3,
\end{equation}
with $\eta_v = 0.21/\pi$, where the assumption is made that in
equilibrium $\dot{E}_{\rm kin} = \dot{E}_{\rm in}$.  The dependence on
the mass of the cube $m$ comes strictly from dimensional arguments, as
all of the runs in Table~\ref{runs} have the same mass $m = \rho_0 L^3
= 8$.  The strong density fluctuations typical of strongly supersonic
turbulence suggest that using the kinetic energy rather than the
volume averaged velocity might give a rather different result.  In
Table~\ref{runs} I give the ratio $E_{\rm kin} / 0.5 m v^2_{\rm rms}$,
showing that in most cases the kinetic energy is 10--15\% higher than
would be expected for perfectly uncorrelated density and velocity
fluctuations.  Fitting to the kinetic energies rather than the
velocities, as shown in Figure~\ref{diss}{\em (b)}, the coefficient
$\eta_e$ is about 20\% different from the equivalent derived from
$\eta_v$, and the slope of the relation actually moves slightly away
from unity to 1.04.  The best linear relation is then
\begin{equation} \label{ediss}
\dot{E}_{\rm kin} = -\eta_e m^{-1/2} \tilde{k} E_{\rm kin}^{3/2}, 
\end{equation}
with $\eta_e = 0.71/\pi$, where the mass dependence is again included on
dimensional grounds.  Equation~(\ref{vdiss}) is not only a slightly
better fit, but it also brings the other hydrodynamic and MHD models
into somewhat better agreement with the relation, so it is mildly
preferred.

The MHD models that fit the relation most closely are the strong field
cases, with $v_A/c_s = 10$.  The weak field cases appear to follow a
relation similar to equation~(\ref{vdiss}), but with values of
$\eta_v$ up to a factor of two higher, as shown in Figure~\ref{beta}.
Without further computation, it remains unclear how much of the
variation seen among the models is due to the remaining lack of
numerical convergence, and how much is real.  The higher dissipation
seen in the high-$\beta$, weak-field cases can be qualitatively
explained by noting that weak fields will be more strongly influenced
by the flow, generating more dissipative MHD waves.  Maron \&
Goldreich (1999) have used a heavily modified version of ZEUS-3D to
compute a relation equivalent to equation~(\ref{ediss}) for strongly
magnetized, trans-Alfv\'enic turbulence, and find a coefficient
equivalent to $\eta_e = 0.23 \pm 0.05$, equal to our value for
hydrodynamic turbulence, and agreeing with our result that the strongly
magnetized models behave very similarly to the hydrodynamic models.

\section{Discussion}

\subsection{Decay Time vs.\ Collapse Time}
An interesting astrophysical question is whether decaying turbulence
can delay gravitational collapse.  We can gain insight into this
question by examining whether the ratio
\begin{equation} \tau = t_d / t_{\rm ff} > 1, \label{tau1} \end{equation} 
where the formal turbulent decay time $t_d = E_{\rm kin} /
\dot{E}_{\rm kin}$, and the free-fall time $t_{\rm ff}$ for the gas is
given by equation~(\ref{tff}).  Because $t_d$ depends not only on the
strength of the turbulence, but also on the driving wavelength, the
value of $\tau$ also depends on the ratio
\begin{equation} \kappa  = \lambda_d / \lambda_J, \label{kappa1} \end{equation}
where the driving wavelength $\lambda_d = 2\pi/\tilde{k}$, and the
Jeans wavelength $\lambda_J = c_s \sqrt{\pi/G \rho_0}$.  It has been
argued that turbulence cannot support the gas against collapse at
wavelengths shorter than the driving wavelength (Bonazzola et al.\
1987, 1992; L\'eorat, Passot, \& Pouquet 1992), so that $\kappa \leq
1$.  This appears likely, but has not yet been confirmed numerically
or observationally.  I will address this issue in future work.

Substituting for the values in equation~(\ref{tau1}), we can write
\begin{equation}
\tau = \frac{E_{\rm kin}}{\dot{E}_{\rm kin}} \frac{c_s}{\lambda_J}
\sqrt{\frac{32}{3}}. 
\end{equation}
We can now use equation~(\ref{vdiss}) for $\dot{E}_{\rm kin}$, and,
somewhat less accurately, take $E_{\rm kin} \sim m v^2_{\rm rms} / 2$,
noting that this introduces no more than a 20--30\% error as shown in
Table~\ref{runs}.  Substituting and using the definition of $\kappa$
given in equation~(\ref{kappa1}), I find that the dissipation time
scaled in units of the free fall time is
\begin{equation}
\tau(\kappa) = \frac{1}{4 \pi \eta_v} \left(\frac{32}{3}\right)^{1/2}
\frac{\kappa}{M_{\rm rms}} \simeq \,3.9 \,\frac{\kappa}{M_{\rm rms}},
\end{equation}
where $M_{\rm rms} = v_{\rm rms}/c_s$ is the rms Mach number of the
turbulence.  In molecular clouds, $M_{\rm rms}$ is typically observed to
be of order 10 or higher.  If $\kappa < 1$ as argued above, then
turbulence will decay long before the cloud collapses and not markedly
influence its collapse.

\subsection{Comparison to Computations of Decaying Turbulence}
\label{decay}
In Paper~I we examined numerical models of decaying supersonic
hydrodynamic and magnetized turbulence and found that its kinetic
energy decayed as
\begin{equation} \label{edecay}
E_{\rm kin}(t) = E_{k0} (1 + t/t_0)^{-\alpha} 
\end{equation}
with $0.85 < \alpha < 1.1$, where $E_{k0}$ is the energy at $t = 0$,
and $t_0$ is a time constant that I will discuss below.  If we
differentiate this, we find
\begin{equation}
\dot{E}_{\rm kin} = - (\alpha / t_0) E_{k0}^{-1/\alpha} E_{\rm
kin}^{1+1/\alpha}. 
\end{equation}
If we compare this to equation~(\ref{ediss}), we see that the energy
would have to decay with a rate $\alpha = 2$ for consistency, rather
than the rate $\alpha \simeq 1$ found in Paper I.

The resolution of this contradiction appears to be that the effective
driving wavenumber $k$ in decaying turbulence is not constant but
decreases over time.  In Figure~\ref{decaymorph} I show cuts through
the density distribution of models C and D of supersonic hydrodynamic
decaying turbulence from Paper I showing a visible increase in the
typical size of structures as time passes.  I show models with both
$128^3$ and $256^3$ resolution to show that the growth in typical size
is not dependent on the resolution, although the detailed structure of
the models certainly is.  To try to demonstrate what such a growth in
typical size ought to look like, I show in Figure~\ref{hydromorph}
slices through models HC2, HC4, HC8, HE2, HE4, and HE8 (see
Table~\ref{runs}).  The MHD case appears more complex.  In
Figure~\ref{decaymhd} I show cuts parallel and perpendicular to the
magnetic field for the $256^3$ decaying model Q from Paper I of
supersonic turbulence in the presence of a strong field with initial
Alfv\'en number unity and initial $\beta = 0.005$.  The length scale
does appear to increase in the structure along the field shown in the
parallel slices, but not in the structure across the field shown in the
perpendicular slices.

In future work I will try to quantify the growth in typical scales
described here, but for now I confine myself to analytically
predicting what the time dependence of the effective driving scale
${\cal L}$ should be.  We can rewrite equation~(\ref{ediss}) in terms
of ${\cal L}(t)$ as
\begin{equation}
\dot{E}_{\rm kin} = 2 \pi \eta_e m^{-1/2} {\cal L}(t)^{-1} E_{\rm kin}^{3/2}, 
\end{equation}
and integrate it, assuming that decay from the driven steady state
begins at $t = 0$ with an equilibrium energy of $E_{k0}$ to find 
\begin{equation} \label{elong}
E_{\rm kin}(t) = E_{k0} \left[1 + \frac{\pi\eta_e E_{k0}^{1/2}}{m^{1/2}} 
  \int_0^t \frac{dt'}{{\cal L}(t')} \right]^{-2}.
\end{equation}
Now we need to find a functional form for ${\cal L}(t)$ that will give
a consistent result.  A useful choice is
\begin{equation} \label{lscale}
{\cal L}(t) = {\cal L}_0 (1 + t/t_0)^{-\alpha/2}, 
\end{equation}
where ${\cal L}_0$ is the driving scale at $t=0$.  Substituting this
into equation~(\ref{elong}) and integrating, we find
\begin{equation}
E_{\rm kin}(t) = E_{k0} \left\{1 + \frac{2 \pi\eta_e E_{k0}^{1/2}}{\alpha
m^{1/2}} \frac{t_0}{{\cal L}_0}  [(1 + t/t_0)^{\alpha/2} - 1]\right\}^{-2}.
\end{equation}
This expression reduces exactly to the empirical form given by
equation~(\ref{edecay}) if and only if the decay time is given by 
\begin{equation}\label{to}
t_0 = \frac{\alpha m^{1/2} {\cal L}_0}{2 \pi \eta_e E_{k0}^{1/2}},
\end{equation}
thus fixing the value of the decay time $t_0$ and showing that my
measured energy dissipation rates in driven turbulence are consistent
with the decay rates measured in Paper~I so long as
equation~(\ref{lscale}) for the effective driving scale holds. 

If we make the assumption again that $E_{k0} \simeq m v^2_{\rm rms}/2$, we
can see that
\begin{equation} \label{t0}
t_0 = \frac{\alpha}{\eta_e \pi \sqrt{2}}\frac{{\cal L}_0}{v_{\rm rms}}.
\end{equation}
Remarkably, the coefficient is empirically found to be unity, as
$\eta_e$ was found in the previous section to be almost exactly
$(\pi\sqrt{2})^{-1}$, and Paper I showed $\alpha \simeq 1$, so the decay time
is just the turbulent crossing time for the driving scale.

Although the decay time $t_0$ derived here appears to have a different
form, it is actually identical to the decay time $t_d = E_{\rm kin} /
\dot{E}_{\rm kin}$ used in the previous subsection if $\alpha = 1$ and
$E_{\rm kin} = E_{k0}$.  This can be seen by substituting for
$\dot{E}_{\rm kin}$ from equation~(\ref{ediss}) and then comparing to
$t_0$ in equation~(\ref{t0}) to find $t_0/t_d = \alpha$.  Thus, the
conclusions drawn there about the ratio of the decay time to the
free-fall time remain valid even if the driving scale is time
dependent as suggested in this subsection.

\acknowledgments I thank E. Zweibel for collaboration on the analysis
of decaying turbulence presented in \S~\ref{decay}, M. Bate,
A. Burkert, R. Klessen, C. McKee, \AA. Nordlund, M. D. Smith,
J. Stone, and E. V\'azquez-Semadeni for useful and interesting
discussions, and J. Maron and P. Goldreich for permission to quote
their result in advance of publication.  Computations presented here
were performed at the Rechenzentrum Garching of the
Max-Planck-Gesellschaft, and at the National Center for Supercomputing
Applications (NCSA).  ZEUS was used by courtesy of the Laboratory for
Computational Astrophysics at the NCSA. This research has made use of
NASA's Astrophysics Data System Abstract Service.

\clearpage

\begin{center} {\large \bf Figure Captions} \end{center}

\figcaption{Energy dissipation rate for models from Table~\ref{runs}
compared to {\em (a)} $k v_{\rm rms}^3$ or {\em (b)} $k E_{\rm
kin}^{3/2}$, where $k = L/\lambda_d$ is the dimensionless wavenumber,
and the size of the cube $L = 2$ for all runs.  The lines have slope
of unity, and are fit to the hydro models HA8 through HE8, yielding
the values for the dissipation coefficients $\eta_v = 0.21$ and
$\eta_e = 0.71$ (see equations~[\ref{vdiss}] and~[\ref{ediss}]).
Hydrodynamical models are indicated by squares, MHD models by
triangles.  \label{diss}}

\figcaption{Dependence of the dissipation coefficient $\eta_v =
\dot{E}_{\rm kin} / (k m v_{\rm rms}^3)$ on the plasma $\beta$, the ratio of
thermal to magnetic pressure.  Models with weaker fields appear to
have as much as a factor two higher dissipation rate. \label{beta}}

\figcaption{Demonstration that typical size scales appear to increase
with time, as suggested by the decay rate. Log of density is shown at
times $t/t_s = 0.2$, 0.6 and 1.0 on slices through the decaying
supersonic hydrodynamic models C and D described in paper I at
standard resolution ($128^3$) and high resolution ($256^3$), where
$t_s$ is the sound crossing time of our numerical box.  Note that each
image is scaled to its own maximum and minimum to enhance
morphological features.
\label{decaymorph}}

\figcaption{Models showing the appearance of turbulence with different
characteristic size scales and driving energy input.  Log of density
on slices through the hydrodynamic models HC2, HC4, HC8, HE2, HE4, and
HE8 (see Table~\ref{runs}) at standard resolution of $128^3$ grid
points.  The value of ``drive'' given in the figure is $\dot{E}_{\rm in}$
for that model.  Note that each image is again scaled to its own
maximum and minimum to enhance morphological features. \label{hydromorph}}

\figcaption{Typical size scales do appear to increase parallel to the
field, but not perpendicular to it.  Log of density is shown in slices
perpendicular and parallel to the magnetic field for the decaying
supersonic, MHD model Q from Paper I, with initial Alfv\'en number
unity and initial $\beta = 0.005$.  Each image is again scaled to its
own maximum and minimum to enhance morphological
features. \label{decaymhd}}

\clearpage
\begin{deluxetable}{lclrlll}
\tablecaption{Uniformly Driven Numerical Models \label{runs}}
\tablehead{
\colhead{Model} &  \colhead{$\dot{E}_{\rm in}$} & \colhead{$k$}  &
\colhead{$v_A/c_s$} & \colhead{$v_{\rm rms}$}  & \colhead{$E_{\rm kin}$}  &
\colhead{$2E_{\rm kin}/mv^2_{\rm rms}$} 
}
\startdata
HA8   & 0.1	       &  8   &	0      & 0.191    &   0.125     & 0.86\nl
HB8   & 0.3	       &  8   &	0      & 0.272    &   0.247     & 0.83\nl
HC2   & 1	       &  2   & 0      & 0.743	  &   2.41$^*$  & 1.1 \nl
HC4   & 1	       &  4   & 0      & 0.530    &   0.961     & 0.86\nl
HC8   & 1	       &  8   &	0      & 0.406    &   0.566     & 0.86\nl
HD8   & 3	       &  8   &	0      & 0.585    &   1.21      & 0.88\nl
HE2   & 10	       &  2   & 0      & 1.50     &   8.92$^*$  & 0.99\nl
HE4   & 10	       &  4   & 0      & 1.19	  &   5.43      & 0.96\nl   
HE8   & 10	       &  8   &	0      & 0.872    &   2.63      & 0.86\nl
      &		       &      &	       &   	  &   	        & \nl
MA4X  & 0.1	       &  4   &	10     & 0.274    &   0.273     & 0.91\nl
MA81  & 0.1 	       &  8   &	1      & 0.147    &   0.0763    & 0.88\nl
MC4X  & 1              &  4   & 10     & 0.534    &   0.787     & 0.69\nl
MC45  & 1	       &  4   &	5      & 0.475    &   0.754     & 0.83\nl
MC41  & 1	       &  4   &	1      & 0.467    &   0.824     & 0.94\nl
MC85  & 1	       &  8   &	5      & 0.335    &   0.351     & 0.78\nl
MC81  & 1	       &  8   &	1      & 0.346    &   0.425     & 0.89\nl
\enddata
\tablecomments{Equilibrium root-mean-square velocities and kinetic energies
of numerical models with $128^3$ numerical resolution.  All quantities
have variances less than five percent, except the two starred kinetic
energies, which have variances of 20\% and 9\%
respectively.}
\end{deluxetable}


\begin{figure}
\figurenum{1}
\plottwo{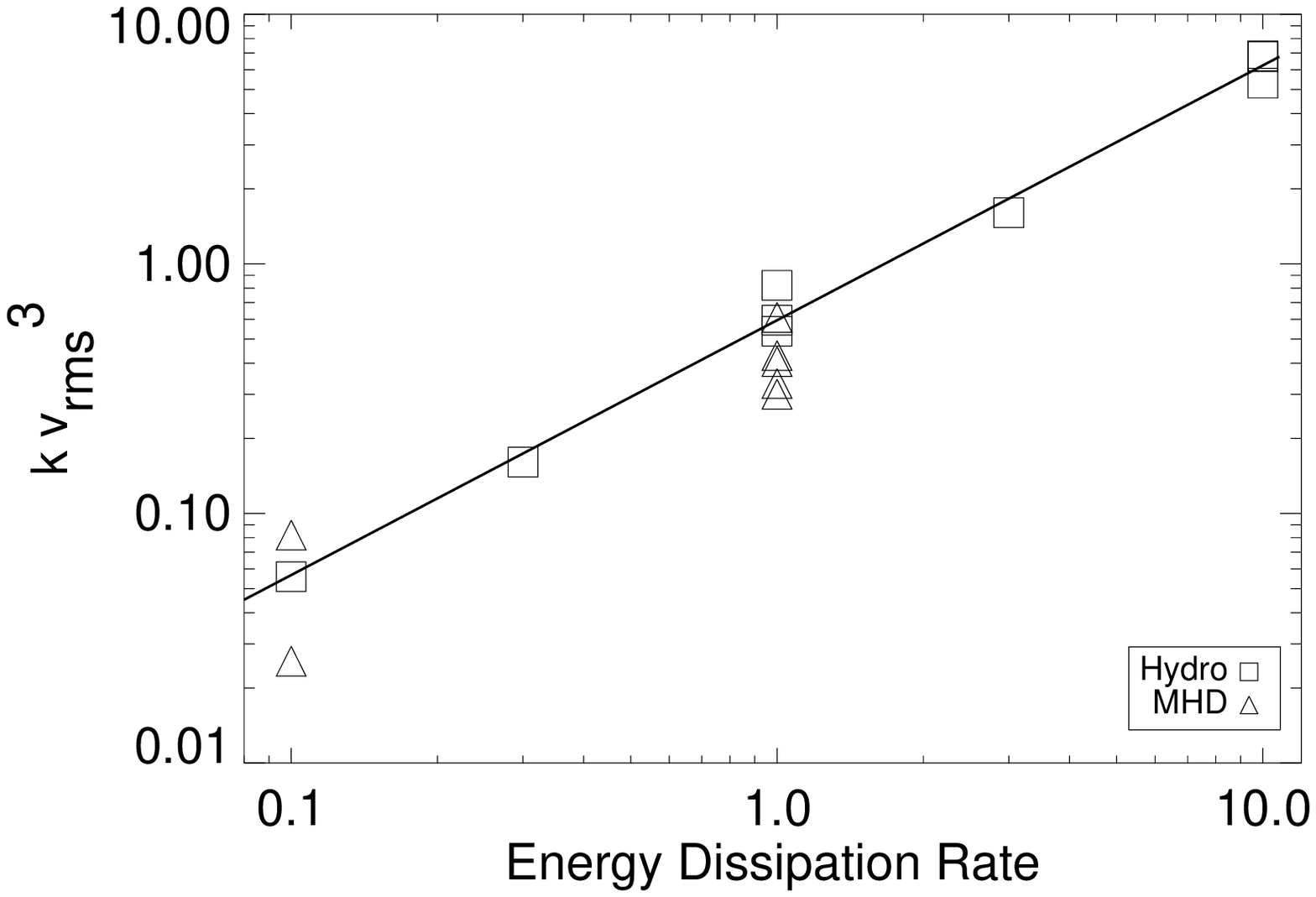}{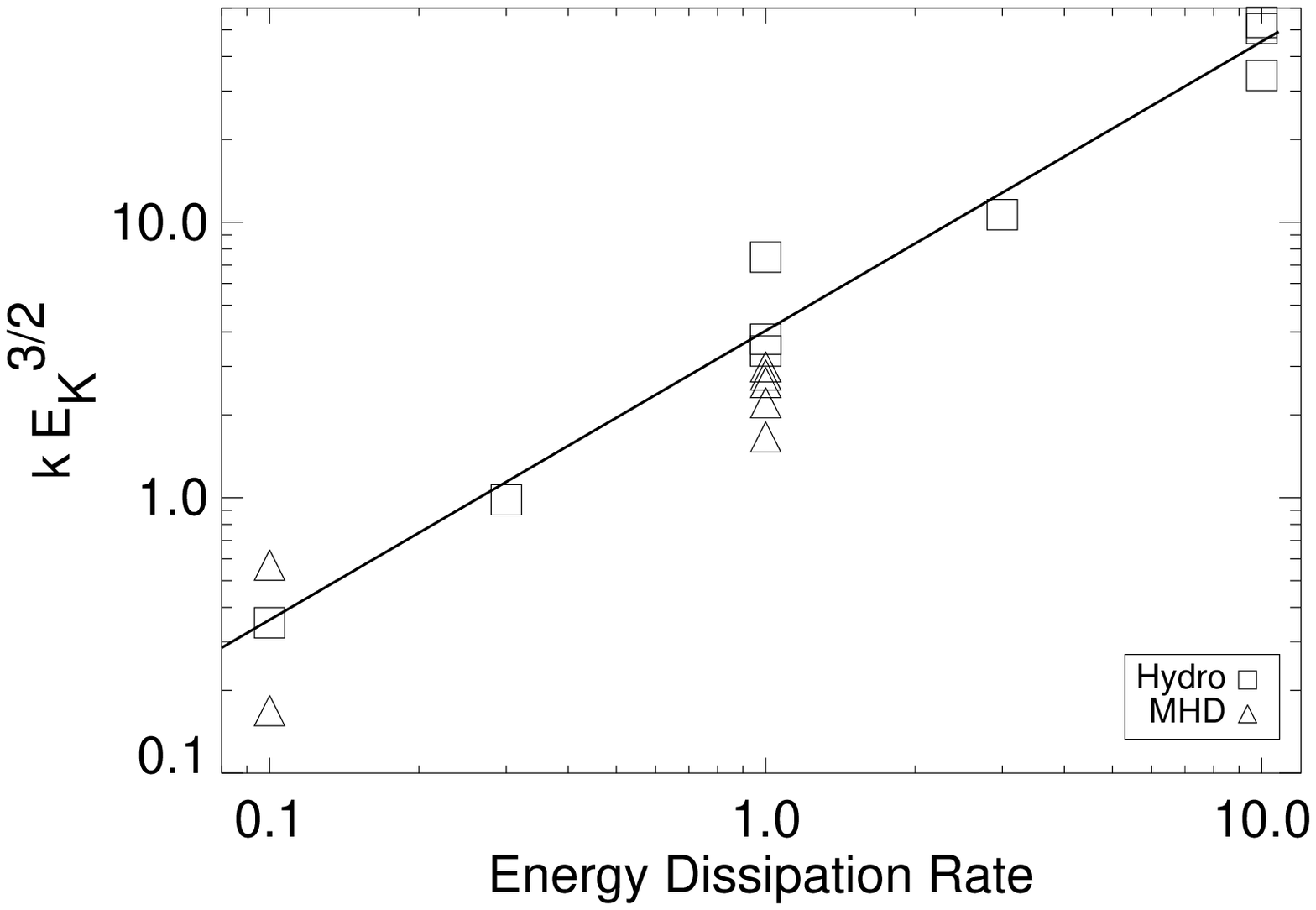}
\caption{Energy dissipation rate for models from Table~\ref{runs}
compared to {\em (a)} $k v_{\rm rms}^3$ or {\em (b)} $k E_{\rm
kin}^{3/2}$, where $k = L/\lambda_d$ is the dimensionless wavenumber,
and the size of the cube $L = 2$ for all runs.  The lines have slope
of unity, and are fit to the hydro models HA8 through HE8, yielding
the values for the dissipation coefficients $\eta_v = 0.21/\pi$ and
$\eta_e = 0.71/\pi$ (see equations~[\ref{vdiss}] and~[\ref{ediss}]).
Hydrodynamical models are indicated by squares, MHD models by
triangles. }
\end{figure}

\begin{figure}
\figurenum{2}
\plotone{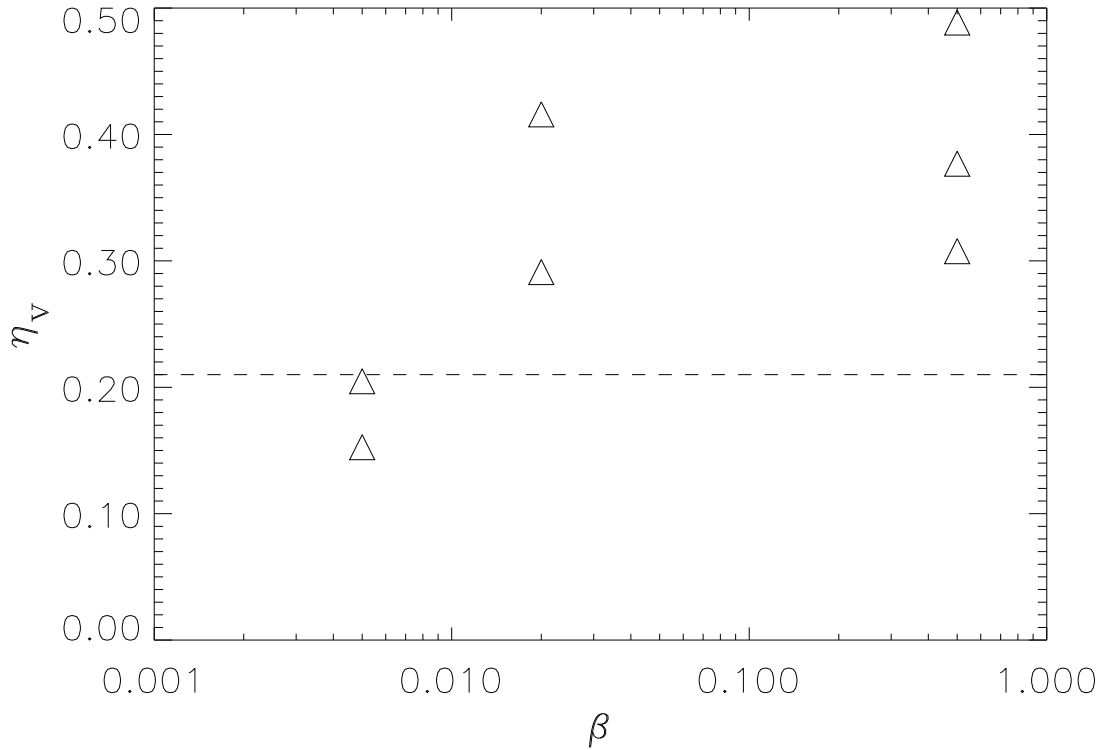}
\caption{Dependence of the dissipation coefficient $\eta_v =
\dot{E}_{\rm kin} / (k m v_{\rm rms}^3)$ on the plasma $\beta$, the ratio of
thermal to magnetic pressure.  Models with weaker fields appear to
have as much as a factor two higher dissipation rate.}
\end{figure}

\begin{figure}
\figurenum{3}
\plotone{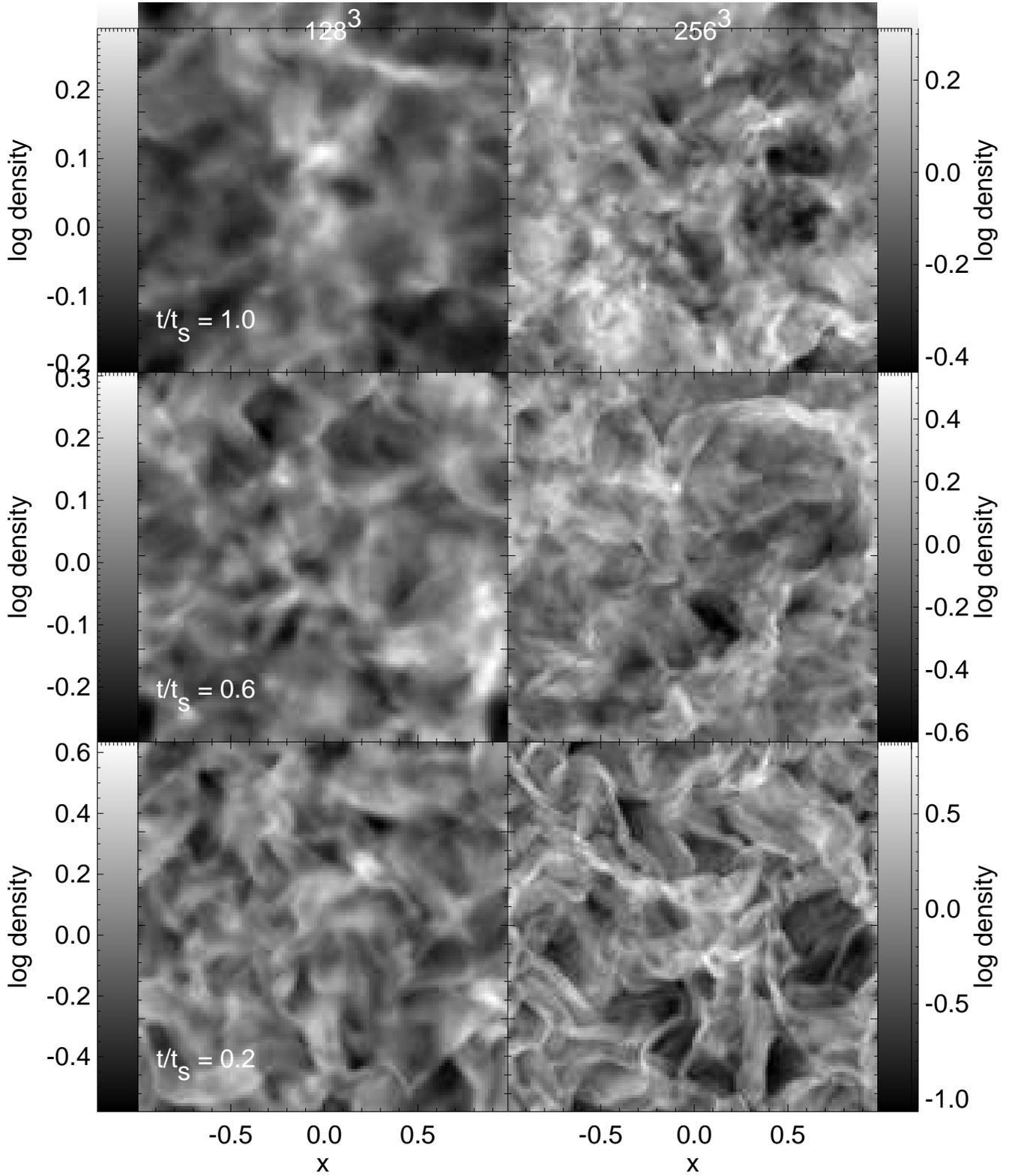}
\caption{Demonstration that typical size scales appear to increase
with time, as suggested by the decay rate. Log of density is shown at
times $t/t_s = 0.2$, 0.6 and 1.0 on slices through the decaying
supersonic hydrodynamic models C and D described in paper I at
standard resolution ($128^3$) and high resolution ($256^3$), where
$t_s$ is the sound crossing time of our numerical box.  Note that each
image is scaled to its own maximum and minimum to enhance
morphological features.}
\end{figure}

\begin{figure}
\figurenum{4}
\plotone{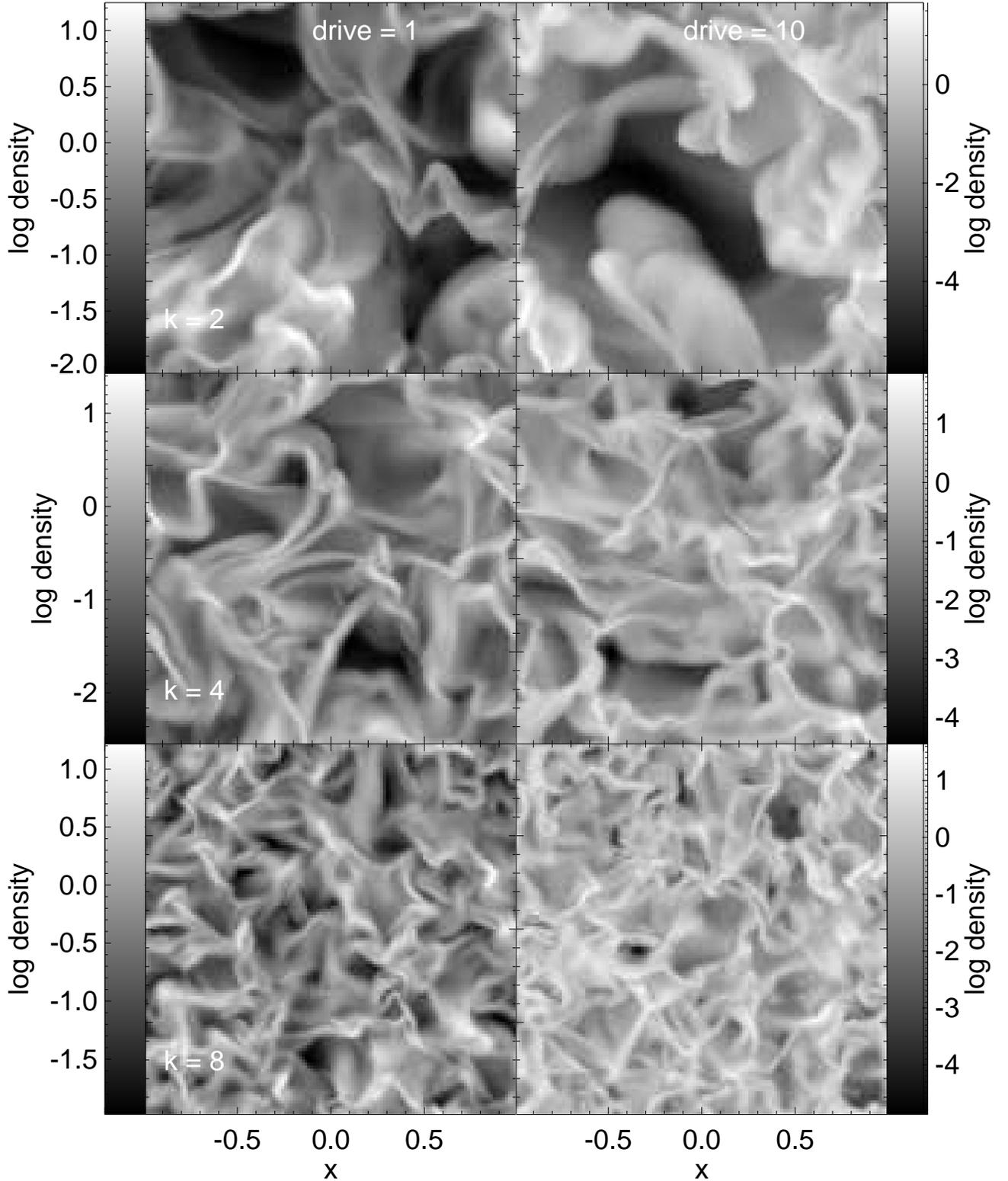}
\caption{Models showing the appearance of turbulence with different
characteristic size scales and driving energy input.  Log of density
on slices through the hydrodynamic models HC2, HC4, HC8, HE2, HE4, and
HE8 (see Table~\ref{runs}) at standard resolution of $128^3$ grid
points.  The value of ``drive'' given in the figure is $\dot{E}_{\rm in}$
for that model.  Note that each image is again scaled to its own
maximum and minimum to enhance morphological features.}
\end{figure}

\begin{figure}
\figurenum{5}
\plotone{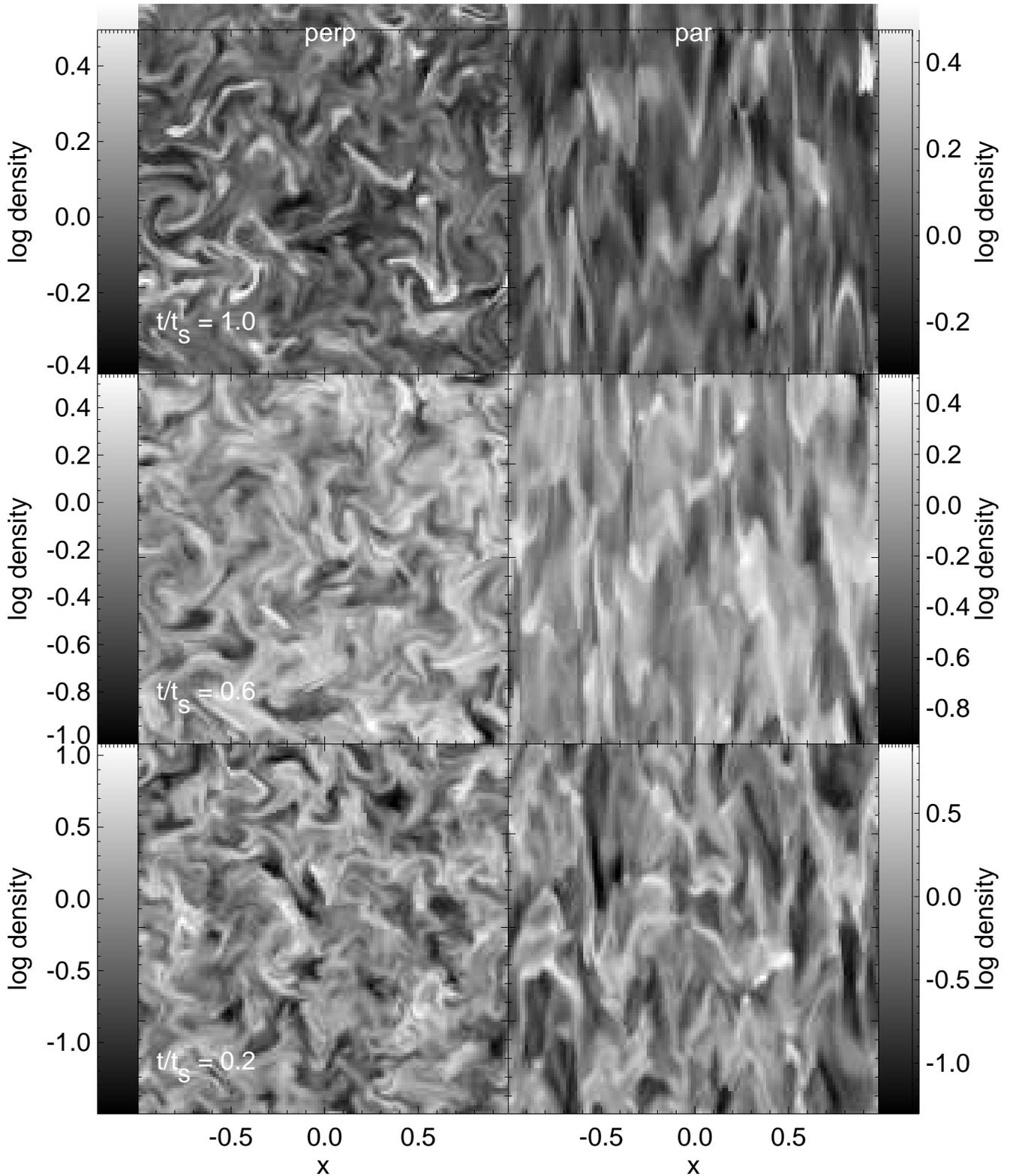}
\caption{Typical size scales do appear to increase parallel to the
field, but not perpendicular to it.  Log of density is shown in slices
perpendicular and parallel to the magnetic field for the decaying
supersonic, MHD model Q from Paper I, with initial Alfv\'en number
unity and initial $\beta = 0.005$.  Each image is again scaled to its
own maximum and minimum to enhance morphological
features.}
\end{figure}

\end{document}